\documentclass{PoS}
\pdfoutput=1

\title{Measurements of Hadron Production at CMS}

\ShortTitle{Hadron Production at CMS}

\author{\speaker{Keith A. Ulmer}\\
        University of Colorado\\
        on behalf of the CMS collaboration\\
        E-mail: \email{keith.ulmer@colorado.edu}}

\abstract{
Measurements of hadron production in $pp$ collisions at $\sqrt{s}$ = 0.9, 2.36 and 7 $\textrm{TeV}$ 
recorded with the CMS detector are reported. Transverse momentum, pseudorapidity and multiplicity 
distributions of charged hadrons are presented. 
For non-single-diffractive collisions, the average 
charged-hadron transverse momentum and pseudorapidity density reveal an increase in production rate 
not well matched by theory and models.
Measured spectra of identified strange particles, 
\PKzS, \PgL, \PagL, \PgXm\ and \PagXp, reconstructed based on their decay topology, 
are also presented. The production rates for strange particles are observed to be in excess
of those predicted by Monte Carlo models by up to a factor of three.}

\FullConference{35th International Conference of High Energy Physics - ICHEP2010,\\
		July 22-28, 2010\\
		Paris France}

\begin{document}

\section{Introduction}

Measurements of particle production are an essential early step in the understanding
of proton-proton collisions at the LHC. Particle spectra provide a testing ground
for the interplay of soft and hard QCD interactions and fragmentation models
at the unprecedented collision energies of the LHC. They also provide valuable
information for the tuning of Monte Carlo (MC) models for future measurements.

CMS is a general purpose experiment at the Large Hadron Collider\,\cite{CMS}.  
The inner detector contains a silicon tracker
composed of pixel layers at radii less than 15\,cm and strip layers out to a radius of
110\,cm.  The central region $|\eta|<1$ has 3 layers of pixels and 10 layers of
strips.  The cylindrical geometry of the central
region changes to disks in the $rz$ plane for the forward region.  Each side of the
interaction region contains two endcap pixel layers and up to 12 layers of strips.
The tracker, PbWO$_4$ electromagnetic calorimeter, and
brass-scintillator hadron calorimeter are all immersed in a 3.8~T axial magnetic field.

The results presented here rely on the excellent performance of the tracking 
system~\cite{trk_10_001} and
come from data collected in 0.9 and 2.36 $\textrm{TeV}$ pp
collisions in December 2009 and from 7 $\textrm{TeV}$ collisions in early 2010.  To preferentially
select non-single-diffractive (NSD) events, activity on both sides of the interaction
region was required using either Beam Scintillation Counters covering $3.23 < |\eta|
< 4.65$ or the forward calorimeter covering $2.9 < |\eta| < 5.2$.  
Beam-halo and beam background events were also removed.

\section{Charged hadron production}
CMS has measured the distribution of charged
particles versus pseudorapidity $(dN_\textrm{ch}/d\eta)$ with three independent methods.
The first method uses counts of pixel clusters, requiring that the cluster shape be
consistent with originating from the interaction region, to measure the number of charged
particles with $p_T$ as low as 30\,MeV/$c$.  The second method uses tracks found in the
pixel system with a minimum $p_T$ of 50\,MeV/$c$.  The third method uses the full silicon
system to reconstruct tracks and is able to reconstruct tracks as low as
$p_T>100\,\textrm{MeV}/c$.  After correcting for
acceptance and efficiency, the three methods give consistent results and are averaged to
obtain the final result\,\cite{dndeta_1,dndeta_2}. 

\begin{figure}[h]
\begin{center}
\includegraphics[clip,height=3.0in]{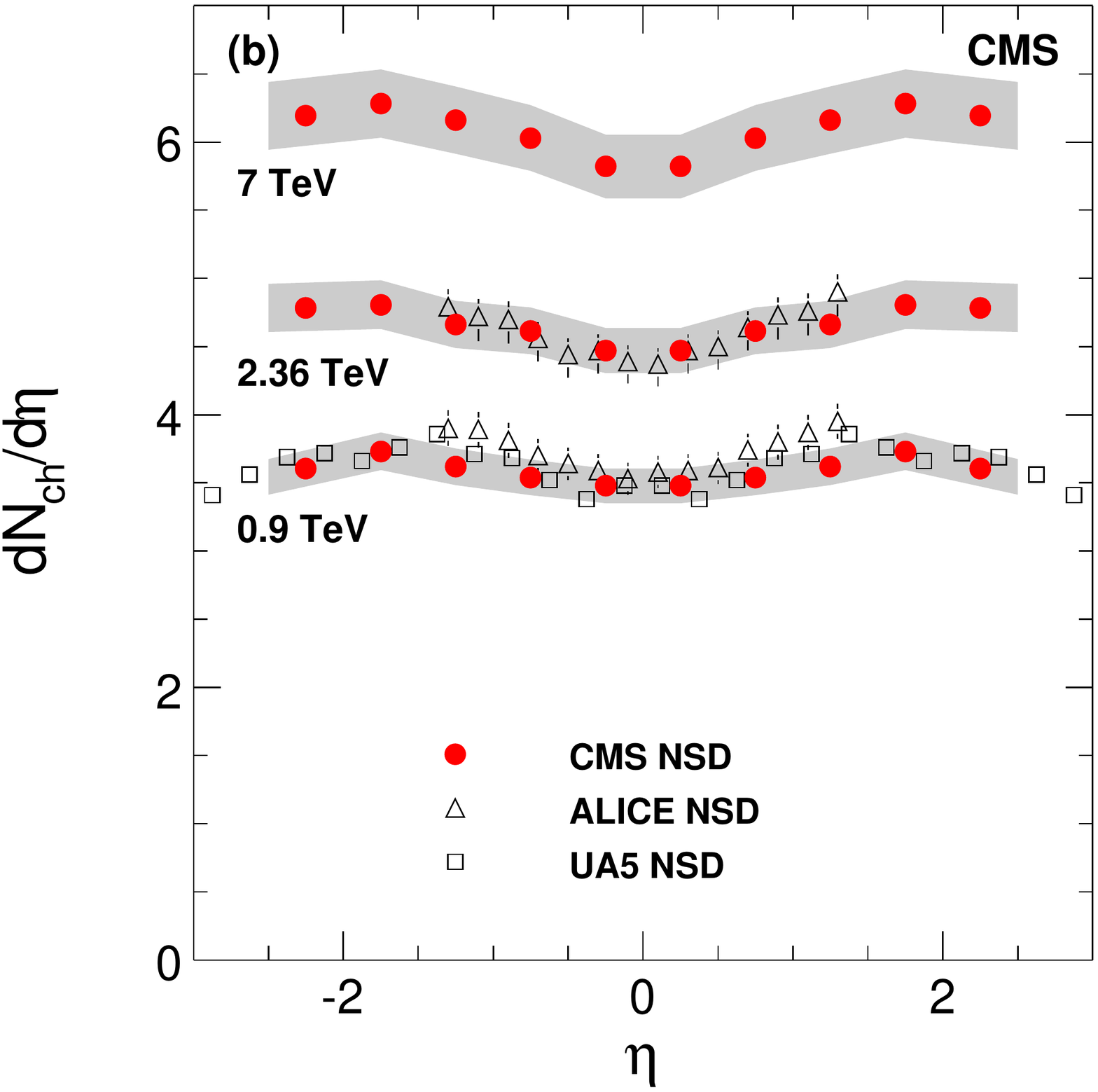}
\includegraphics[clip,height=3.0in]{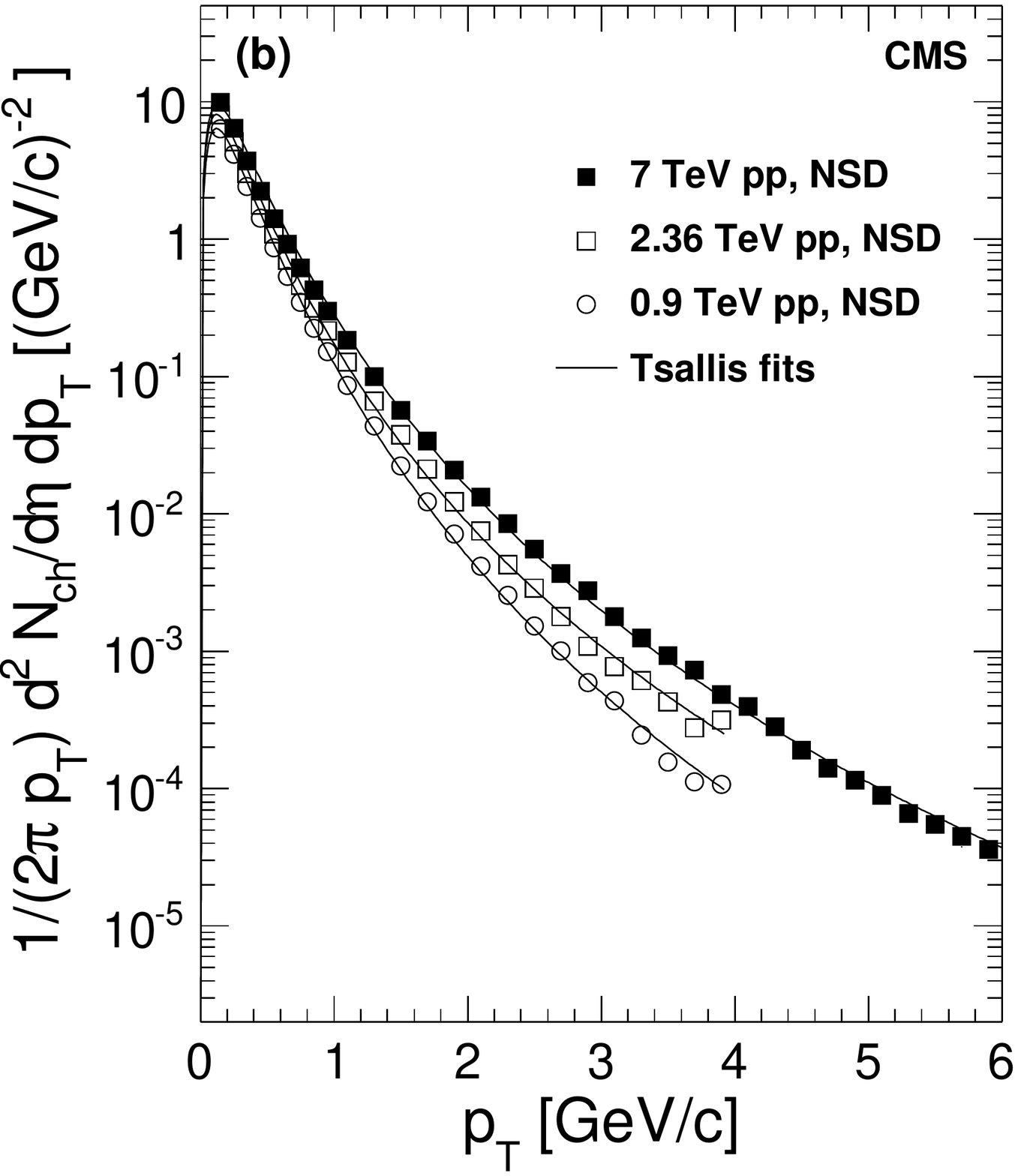}
\caption{$dN_\textrm{ch}/d\eta$ versus $\eta$ (left) and 
$dN_\textrm{ch}/d p_T$ versus $p_T$ (right).
The $\eta$ results include previous results from UA5 and ALICE\@.  
The shaded bands show the systematic uncertainty of the CMS result.}
\label{fig:dndeta}
\end{center}
\end{figure}

Figure~\ref{fig:dndeta} shows the distribution
$dN_\textrm{ch}/d\eta$ for NSD events for $\sqrt{s}$ = 0.9, 2.36 and 7 $\textrm{TeV}$. 
The results are shown to be in good agreement with measurements from other experiments.
The full tracking method also provides a measurement of track $p_T$. The distributions 
$dN_\textrm{ch}/d p_T$ for NSD events are also shown in Fig.~\ref{fig:dndeta} and 
demonstrate the characteristic sharply falling distribution as a function of $p_T$.

\begin{figure}[h]
\begin{center}
\includegraphics[clip,height=3.0in]{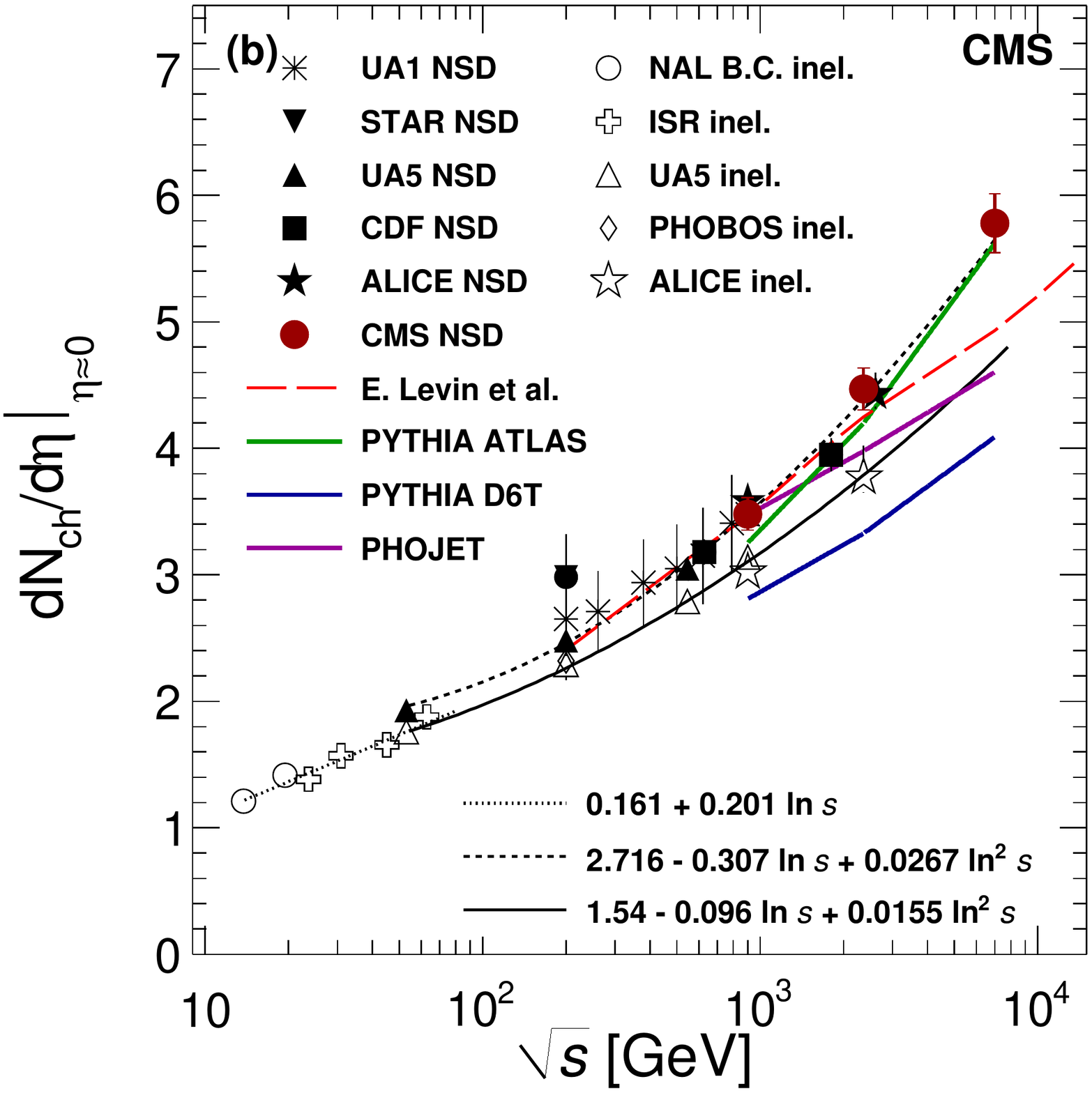}
\includegraphics[clip,height=3.0in]{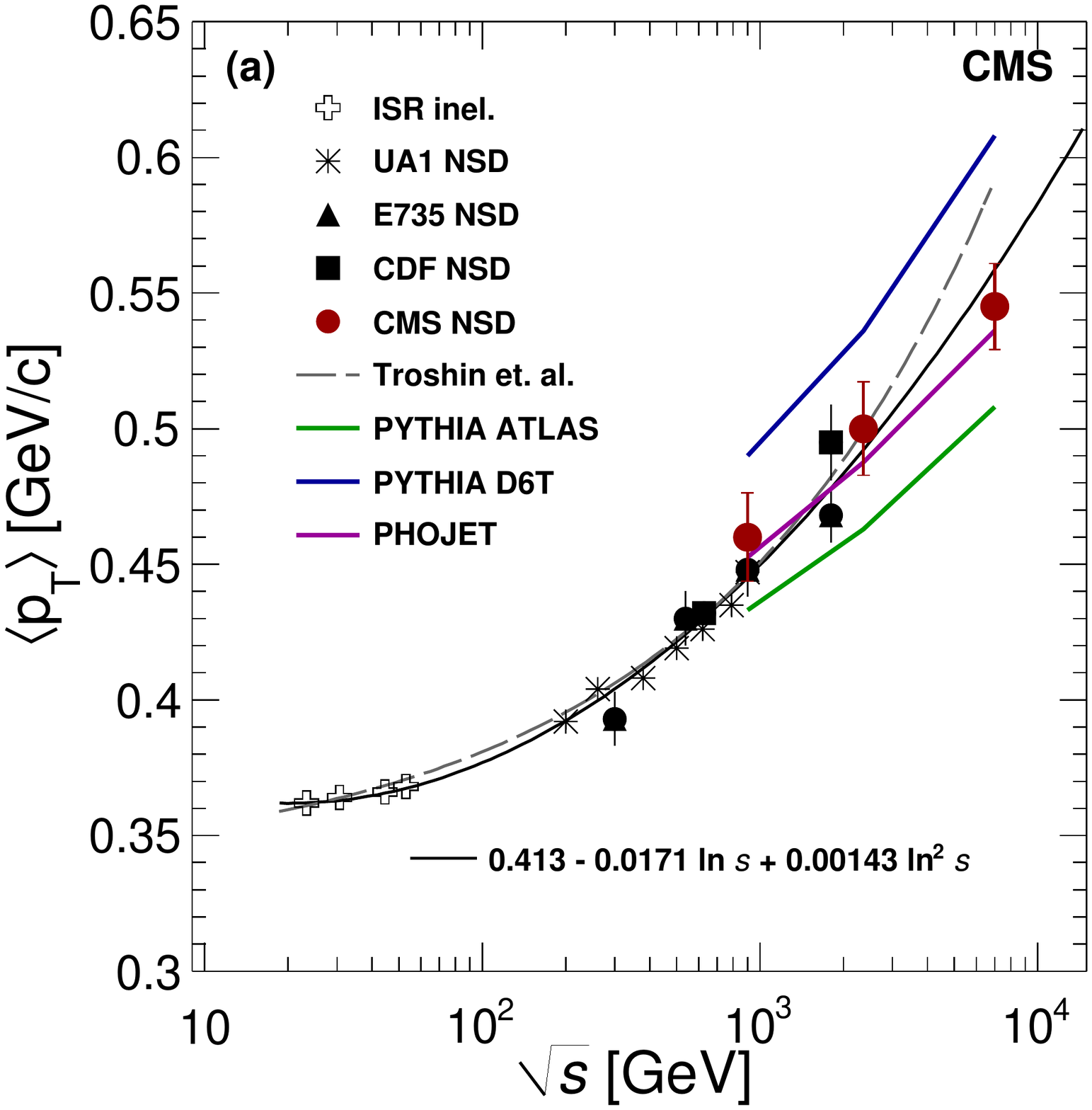}
\caption{$dN_\textrm{ch}/d\eta$ at $\eta \approx 0$ (left) and
average $p_T$ of charged tracks (right) versus $\sqrt{s}$. Results from CMS
are shown along with other comparable results and various theoretical
predictions and models.}
\label{fig:dndeta2}
\end{center}
\end{figure}

The observed increase in multiplicity and $p_T$ 
versus center-of-mass energy is seen clearly in
Fig.~\ref{fig:dndeta2} where $dN_\textrm{ch}/d\eta$ at $\eta \approx 0$ (left) and
average $p_T$ (right) are plotted versus $\sqrt{s}$.  
While parametrizations of the behavior of
$dN/d\eta|_{\eta\approx 0}$ and $\langle p_T \rangle$ versus $\sqrt{s}$ are possible, the
current \textsc{Phojet}~\cite{Engel:1994vs, Engel:1995yda} and 
\textsc{Pythia}~\cite{Sjostrand:2006za} curves are unable to account for both results.

The charged particle multiplicity distribution
for non-single-diffractive events has also been measured~\cite{QCD-10-004}.
The data are corrected for the trigger and event selection efficiency and the
effects of tracking inefficiency and secondary tracks originating
from the decay of long lived particles and products of interactions with the
beam pipe and the detector material. 

\begin{figure}
\begin{center}
\includegraphics[clip,height=1.7in]{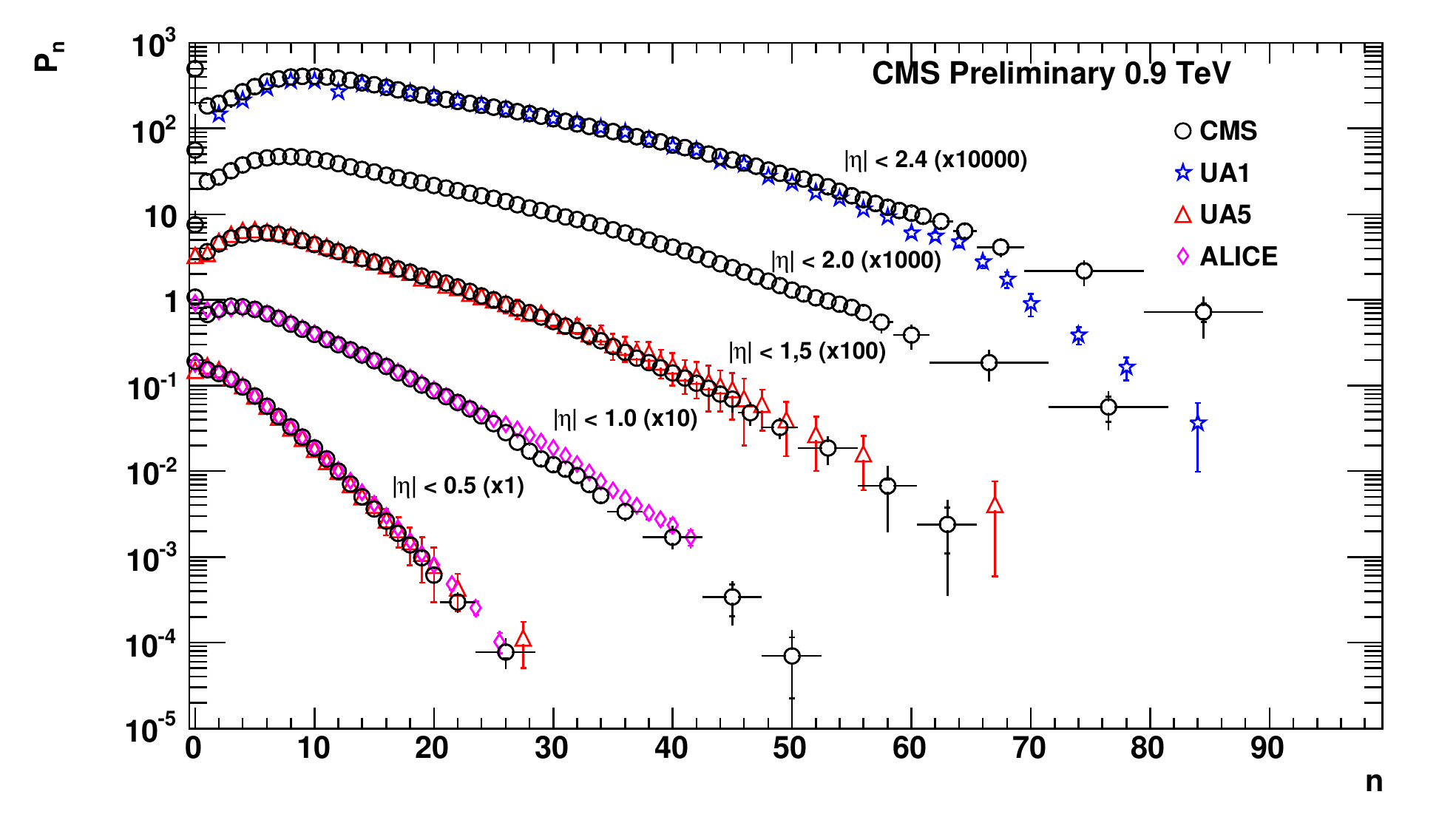}
\includegraphics[clip,height=1.7in]{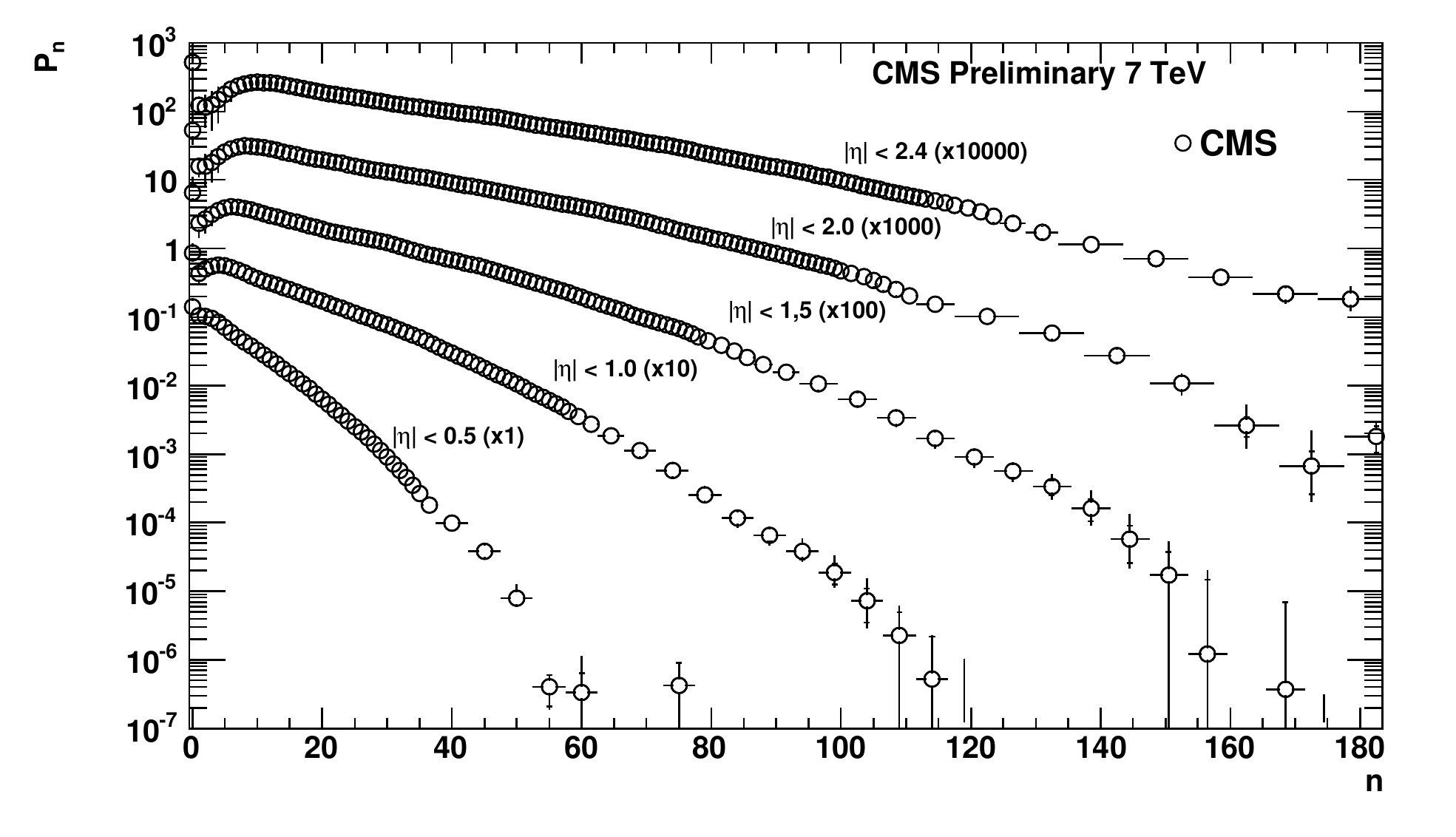}
\caption{Charged particle multiplicity distributions for $\sqrt{s} = 0.9$ (left)
and 7 $\textrm{TeV}$ (right), with comparisons to other measurements for the lower
energy.}
\label{fig:mult}
\end{center}
\end{figure}

Figure~\ref{fig:mult} shows the multiplicity distributions for $\sqrt{s} = 0.9$
and 7 $\textrm{TeV}$ for various domains of pseudorapidity up to $| \eta | < 2.4$.
The multiplicity distribution at $\sqrt{s} = 0.9$ $\textrm{TeV}$ is in
agreement with previous experiments. At higher energies, a strong increase of the mean
multiplicity with $\sqrt{s}$ is observed, which is underestimated by most Monte Carlo models.
The measurement of higher order moments of the multiplicity
distribution confirms the violation of KNO scaling that has been observed at lower
energies.

\section{Strange particle production}

\begin{figure}
\begin{center}
\includegraphics[clip,height=1.8in]{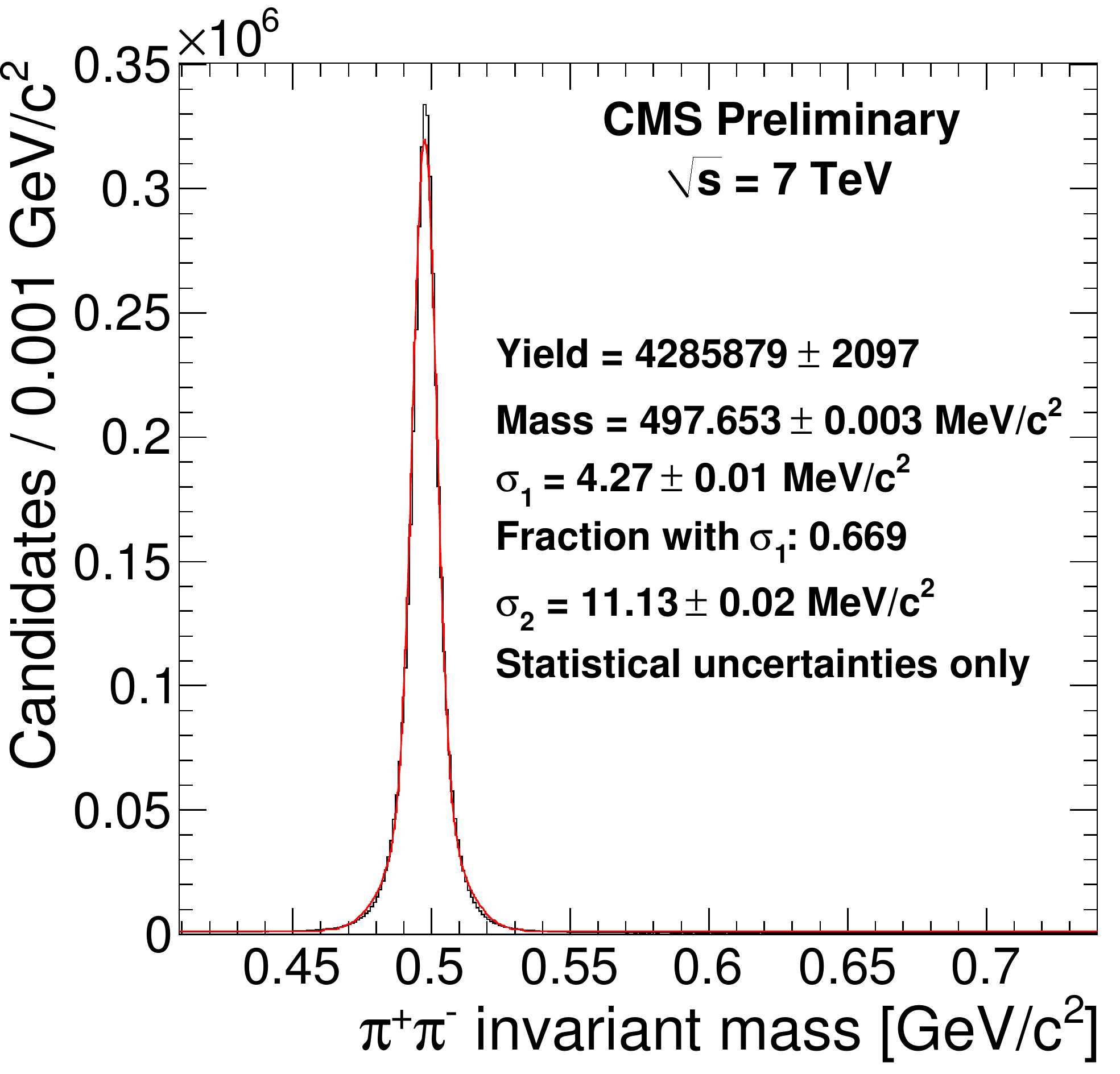}
\includegraphics[clip,height=1.8in]{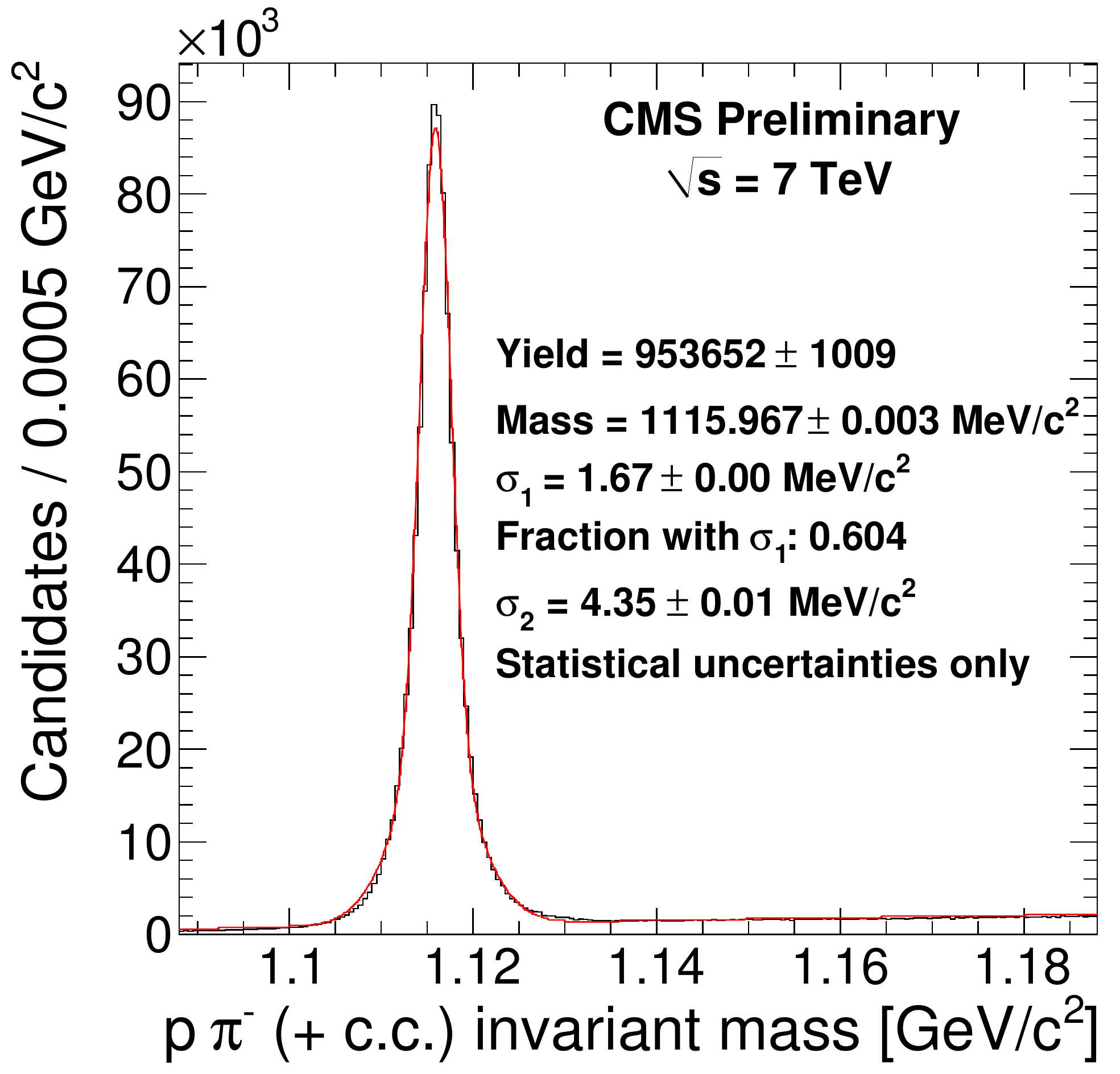}
\includegraphics[clip,height=1.8in]{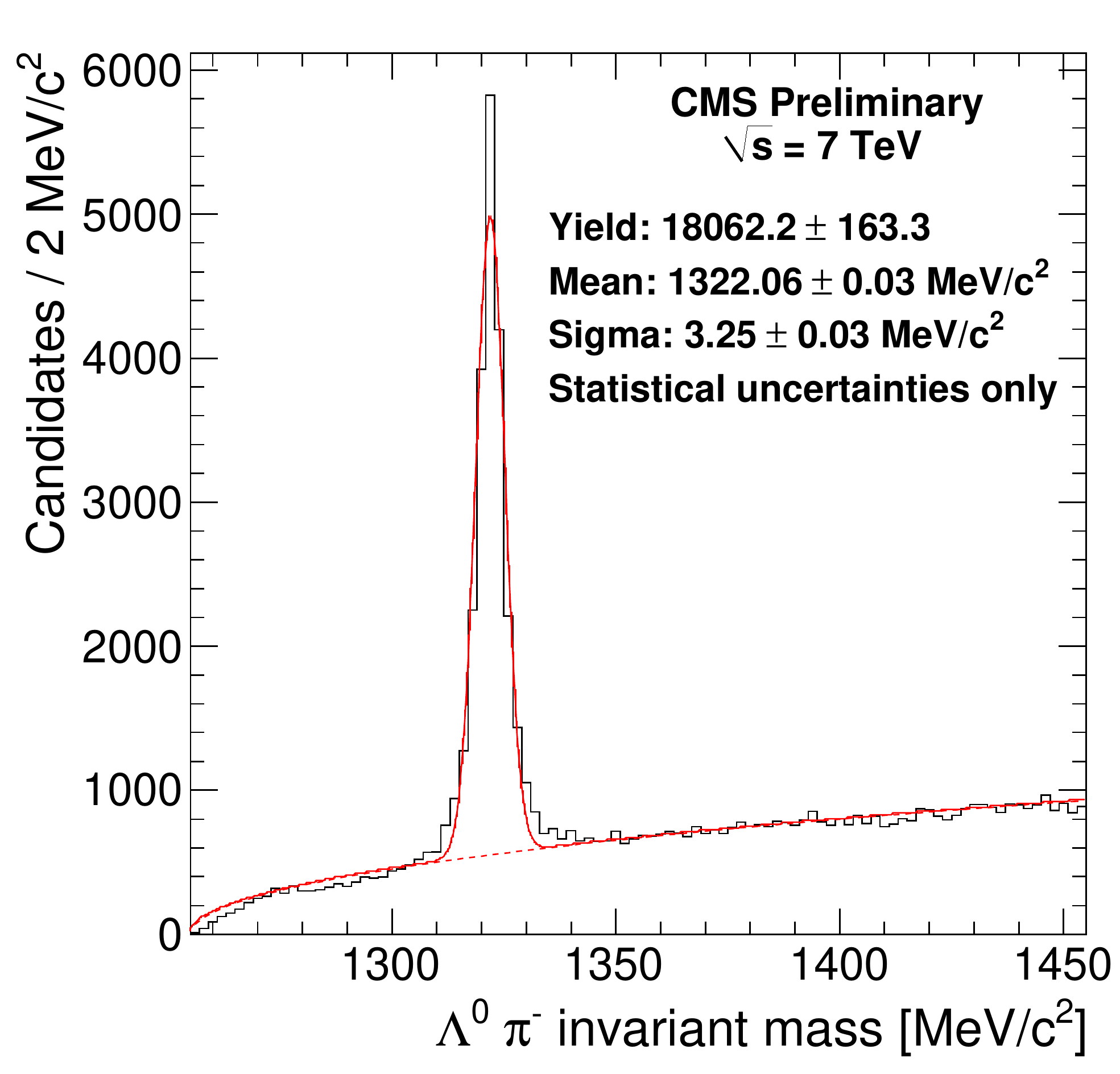}
\caption{Invariant mass distributions for, from left to right, reconstructed 
\PKzS, \PgL\ and \PgXm\ particles from 7 $\textrm{TeV}$ collision data.}
\label{fig:v0masses}
\end{center}
\end{figure}

Long-lived, strange particles, \PKzS, \PgL, \PagL, \PgXm\ and \PagXp, have
been reconstructed and identified based on their decay topology in decays to the
final states $\pi^+\pi^-$, $p\pi^-$ and $\PgL\pi^-$, respectively~\cite{QCD-10-007}, 
with charge conjugates implied throughout. 
\PKzS\ and
\PgL\ particles are reconstructed by fitting oppositely charged tracks that
are displaced from the primary interaction vertex to a
common secondary vertex that is also required to be displaced from the primary. 
\PgXm\ particles are reconstructed by combining a \PgL\ candidate with a third displaced
charged track. Figure~\ref{fig:v0masses} shows the reconstructed invariant mass 
distributions for \PKzS, \PgL\ and \PgXm\ from 7 $\textrm{TeV}$ data.

\begin{figure}
\begin{center}
\includegraphics[clip,height=1.8in]{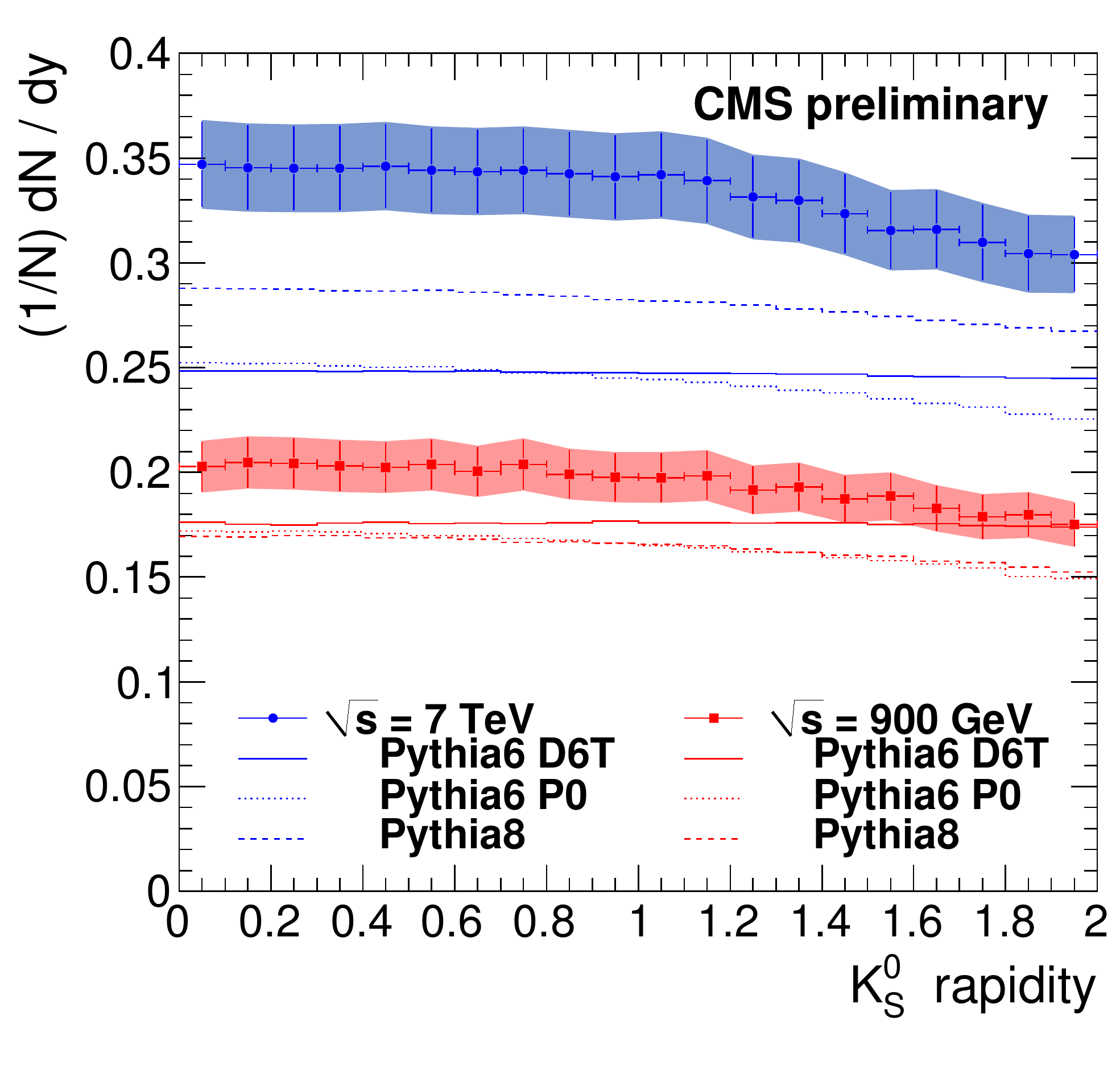}
\includegraphics[clip,height=1.8in]{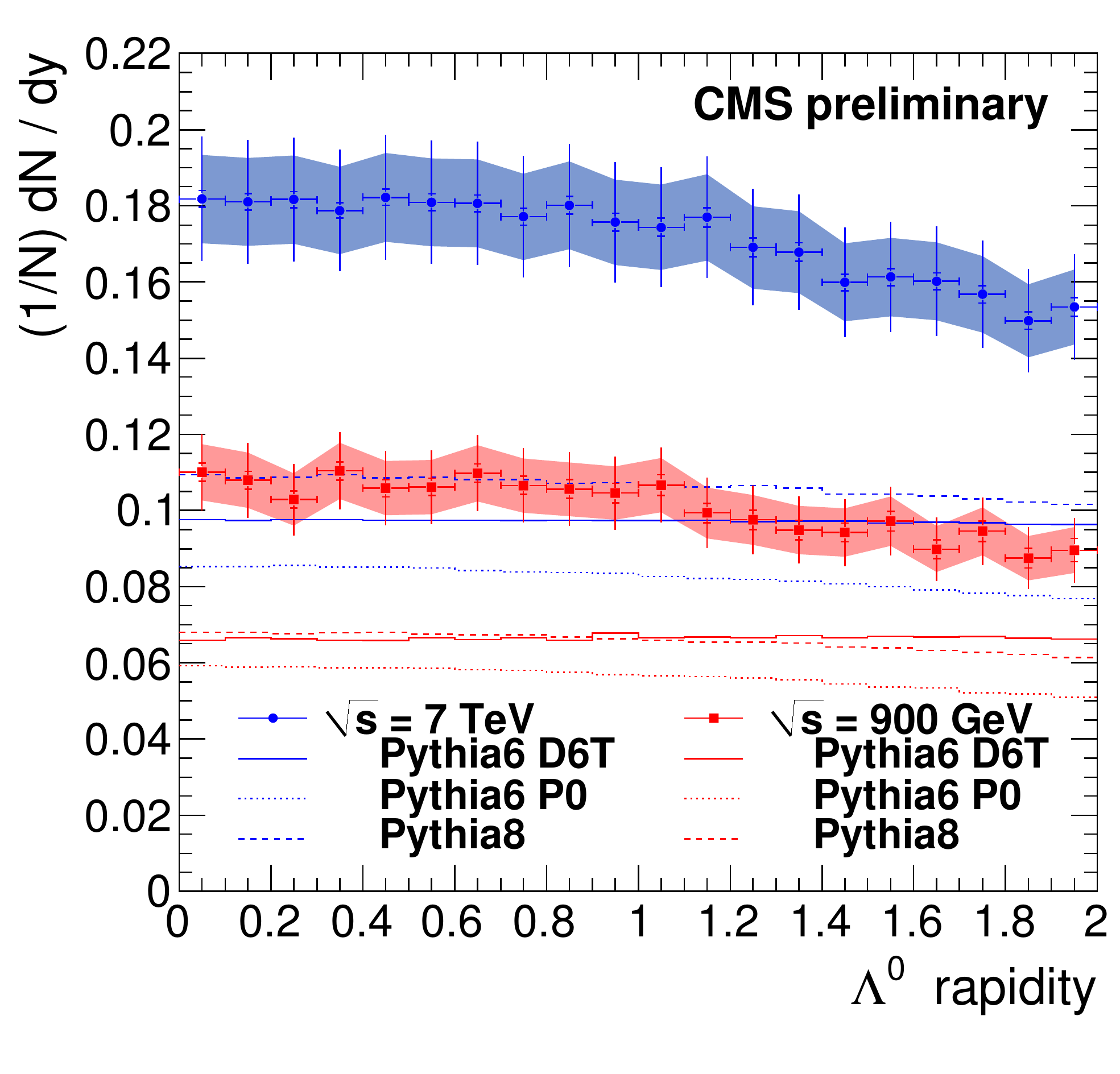}
\includegraphics[clip,height=1.8in]{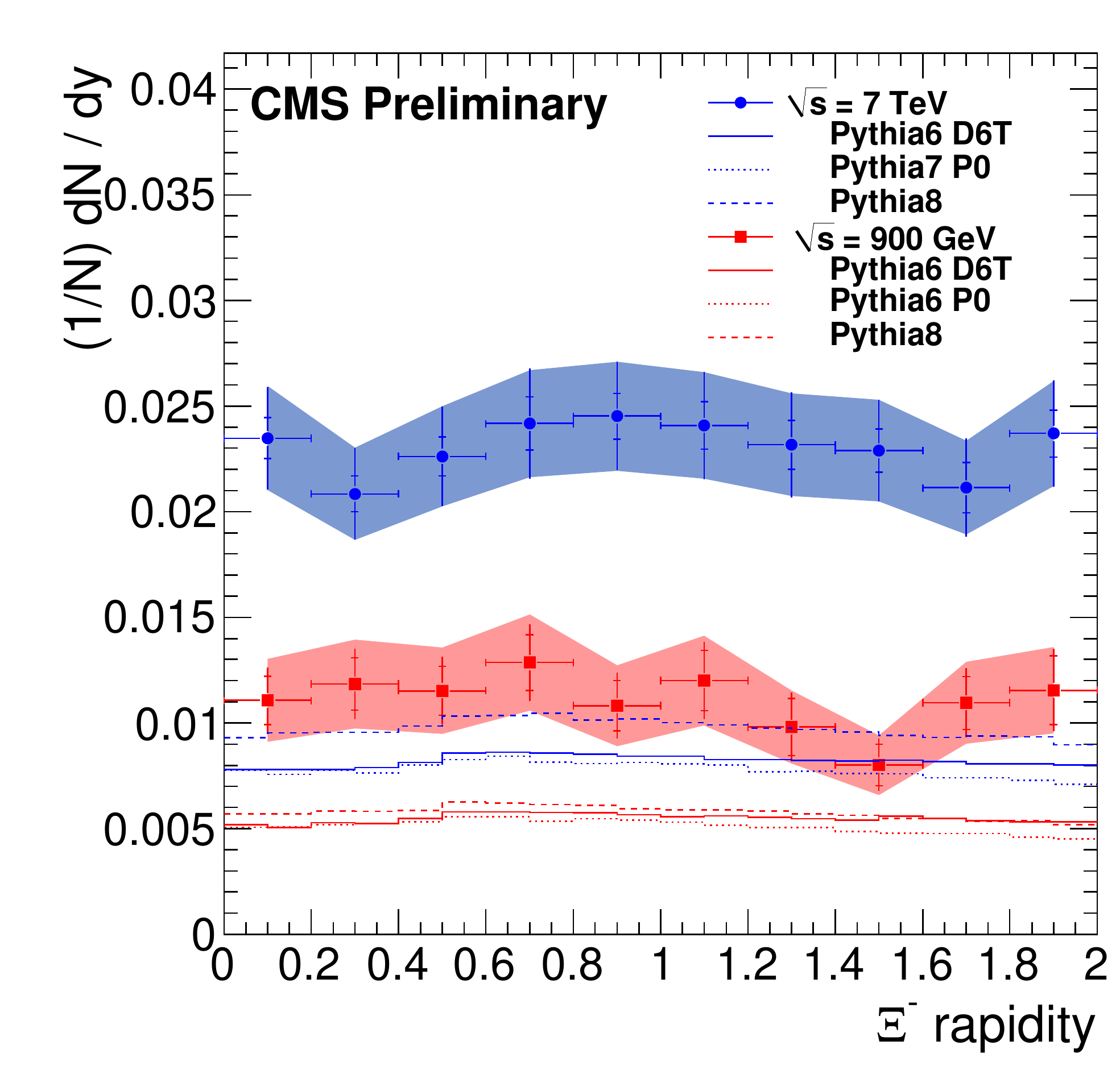}
\caption{$dN/dy$ distributions for, from left to right, \PKzS, \PgL\ and 
\PgXm\ particles. Predictions from various $\textsc{Pythia}$ tunes
are also shown.}
\label{fig:v0results}
\end{center}
\end{figure}

The reconstruction efficiency for each particle species is determined from MC 
simulation and is used to scale the reconstructed particle yields to measure the
production distributions as functions of rapidity, $y$, and $p_T$. The MC is observed to
be a poor representation of the data in both track multiplicity and $p_T$ and $y$ for the 
strange particles and is reweighted to match the data. The particle yields are corrected
for trigger and event selection efficiency, non-prompt and single
diffractive contributions to measure $dN/dy$ and $dN/dp_T$ for NSD events for each
particle species. The measured distributions for $dN/dy$ are shown in 
Fig.~\ref{fig:v0results}. The expected production rates from various $\textsc{Pythia}$ models
are also shown, and are found to be significantly lower than measured in the data, up
to a deficit of a factor of three for \PgXm\ production at 7 $\textrm{TeV}$.

\section{Conclusion}

Charged track and strange particle production rates have been measured as
functions of transverse momentum and (pseudo)rapidity in CMS data at 
$\sqrt{s}$ values of 0.9, 2.36 and 7.0 $\textrm{TeV}$. The rise in charged
particle production rate with $\sqrt{s}$ is not well modeled, nor is the 
overall rate of strange particle production, which is observed to exceed
predictions from $\textsc{Pythia}$ by up to a factor of three.


\begin{thebibliography}{99}

\bibitem{CMS}
  The CMS Collaboration,
  \textit{JINST} {\bf 3}, S08004 (2008).

\bibitem{trk_10_001} 
  The CMS Collaboration, arXiv:1007.1988.
  
\bibitem{dndeta_1}
  The CMS Collaboration,
  \textit{JHEP} {\bf 1002}, 041 (2010), arXiv:1002.0621.

\bibitem{dndeta_2}
  The CMS Collaboration, 
  \textit{Phys. Rev. Lett.} {\bf 105} 022002 (2010), arXiv:1005.3299.

\bibitem{Engel:1994vs}
Engel, R., \textit{Z. Phys.} {\bf C66} 203-214 (1995).

\bibitem{Engel:1995yda}
Engel, R. and Ranft, J., \textit{Phys. Rev.} {\bf D54} 4244-4262 (1996),arXiv:hep-ph/9509373

\bibitem{Sjostrand:2006za}
Sjostrand, Torbjorn and Mrenna, Stephen and Skands, Peter Z.,
\textit{JHEP} {\bf 05}, 026 (2006), arXiv:hep-ph/0603175.

\bibitem{QCD-10-004}
The CMS Collaboration, CMS Physics Analysis Summary {\bf QCD-10-004},
http://cdsweb.cern.ch/record/1279343, (2010).

\bibitem{QCD-10-007} 
The CMS Collaboration, CMS Physics Analysis Summary {\bf QCD-10-007},
http://cdsweb.cern.ch/record/1279344, (2010).

\end{thebibliography}
\end{document}